# Does the Forced Van der Pol Oscillator Exhibit Irregular Behavior?

Samaira Tibrewal[1] and Soumyajit Seth[2]

[1] NPS International School, Singapore
[2] School of Technology Management and Engineering, NMIMS Hyderabad, Jadcherla Campus, Polepally, 509301 India

## SUMMARY

In various physical and engineering systems in nature, we encounter different types of oscillators, many of which exhibit nonlinear behavior. One such system, the Van der Pol oscillator, is known for its self-sustained limit cycle behavior. However, when subjected to external forcing, its dynamics can change abruptly from its self-sustained oscillation. This study investigates the onset of complex behavior in the sinusoidally forced Van der Pol oscillator, aiming to understand how the system transitions from regular to aperiodic dynamics under varying conditions. The central research question is: Does the forced Van der Pol oscillator exhibit irregular dynamics such as quasi-periodicity or chaos, and if so, what are the possible conditions?

We hypothesize that if we choose different values of the amplitude and ratios of the external forcing frequency with respect to the natural frequency of the oscillator, it can induce a range of dynamic regimes, including higher periodic and chaotic ones. Using numerical simulations in Python, we analyze the behavior of the system through time series plots, phase portraits, and bifurcation diagrams. Our results reveal that as the bifurcation parameter, the forcing amplitude is varied, having different frequency ratios, the system transitions through periodic, quasi-periodic, and chaotic states. This type of nonlinear interaction has wide applications in real-world systems, including biological rhythms, electrical circuits, and astrophysical processes.

## INTRODUCTION

A Simple Harmonic Oscillator (SHO) exhibits periodic motion where acceleration is directed toward a fixed equilibrium point and is proportional to displacement, resulting in a linear relationship. Applications of SHO include clocks, musical instruments, and metronomes. However, many real-world systems display nonlinear behavior, such as gravitational forces in astrophysics, heartbeats in biology, predator-prey dynamics in ecology, and climate cycles like El Niño and La Niña. Nonlinearity is often introduced into the SHO through damping—a force that dissipates energy, typically as heat, leading to a gradual reduction in amplitude. In systems like pendulums, damping arises from resistive forces such as air resistance or friction, modeled by a velocity-dependent term (e.g., $-b\theta$, where b is the damping coefficient and θ is the angular velocity). This incorporation of damping makes the model more realistic by accounting for energy losses inherent in practical oscillating systems.

The Van der Pol oscillator [1] is a simple nonlinear system that exhibits damped harmonic motion, with damping modeled by a nonlinear velocity-dependent term. This nonlinearity regulates the system's energy, enabling self-sustained oscillations known as a *limit cycle*. The governing equation for the Van der Pol oscillator is: $d^2x/dt^2 + (x^2 − 1)dx/dt + x = 0$

When the displacement x is less than one, the damping term of the Van der Pol oscillator is negative, reducing resistance and increasing velocity. When x exceeds one, the damping becomes positive, increasing resistance and reducing velocity. This self-regulating mechanism drives the system toward a stable limit cycle [2], resulting in periodic motion independent of initial conditions. Manu [3] investigated the existence of limit cycles and the role of the parameter μ in system stability. Tsatsos [4] and Chughtai [5] explored how variations in system parameters—such as damping, frequency, and amplitude—affect transitions between periodic, quasi-periodic, and chaotic behaviors, using *bifurcation diagrams* and *Fourier power spectra*. Venkatesan and Lakshmanan [6] investigated a driven double-well Duffing–Van der Pol oscillator, identifying various attractors and transitions between stable and chaotic behavior, including local-global bifurcations of intermittent and blue-sky catastrophe types. Jiang and Wei [7] studied the Van der Pol oscillator with delayed feedback, revealing transcritical and pitchfork bifurcations and changes in stability associated with a simple zero eigenvalue.

A Van der Pol oscillator becomes forced when subjected to an external periodic force, with its dynamics governed by parameters such as nonlinear damping and the frequency and amplitude of forcing. Depending on these parameters, the system can exhibit synchronization, amplitude modulation, *quasiperiodicity*, or *chaos*, as shown by Ku and Sun [8] and Chughtai [5]. Pinto and Machado [9] extended this study to higher-order forced Van der Pol oscillators, uncovering diverse oscillatory patterns and stability conditions. Bifurcation analyses by Sharma [10] and Arrowsmith and Taha [11] revealed *Saddle-node, Supercritical Hopf,* and *Infinite-period bifurcations*, noting that system symmetry simplifies bifurcation structures and transitions between oscillatory and non-oscillatory states. Xu and Jiang [12] and Holmes and Rand [13] further examined bifurcation behavior under sinusoidal and periodic forcing, observing a range of periodic and aperiodic responses. Ševčík and Přibylová [14] identified the coexistence of a stable torus and a stable cycle emerging from a q-fold bifurcation within an Arnold tongue. Cooper et al. [15] also applied perturbation methods and numerical simulations to control or stabilize chaotic dynamics in the forced Van der Pol oscillator.

As mentioned above, the studies demonstrate that varying the amplitudes of the forced Van der Pol oscillator can produce inherent complex phenomena, including chaotic orbits and bifurcation. In this context, chaos refers to an irregular dynamics, where small changes in initial conditions lead to significantly different long-term outcomes, a property known as *sensitive dependence on initial conditions*. The chaotic behavior arises due to the mismatch of the frequency ratios or the high values of the forcing amplitudes, which makes the oscillation unpredictable.

An open question is how the dynamics of the forced Van der Pol oscillator change as the forcing amplitude varies smoothly, particularly for integer and non-integer ratios of forcing to natural frequency. This study investigates how the dynamical behavior of the system and bifurcation structure evolve under these conditions.

In real-world scenarios, the amplitude of the forcing term often varies. In structural engineering, variable forces from seismic events, wind, or traffic loads influence the dynamics of buildings and bridges. Instruments like guitars and violins exhibit force variations in acoustics due to changes in playing techniques. Similarly, wave-induced forces on ships or offshore platforms fluctuate in fluid-structure interactions with environmental conditions such as tides and storms. These variations significantly impact the oscillatory response of the system.

System Equations: The equation of the forced Van der Pol oscillator is as below [7]:

$$\frac{d^2x}{dt^2} + \mu(x^2 - 1)\frac{dx}{dt} + x = F\cos(\omega t)$$

In the system equation, x denotes displacement, μ is the nonlinear damping coefficient, ω is the angular frequency, and F is the forcing amplitude. This form models various real-world systems—mechanical (e.g., mass-spring systems), electrical (e.g., RLC circuits with AC sources), and biological (e.g., periodic forces from cardiac or respiratory cycles)—via a forcing term Fcos(ωt). When μ=0 and F=0, the equation reduces to the classical Van der Pol oscillator. This study explores the dynamical behavior of the system by varying the forcing amplitude F.

**RESULTS**

To understand how varying the forcing amplitude F influences the dynamics of the forced Van der Pol oscillator, we performed a numerical bifurcation analysis with fixed parameters μ=1 and ω=1. The state variable was sampled stroboscopically, synchronized with the external periodic forcing, to capture the long-term behavior of the system.

The bifurcation diagram (Figure 1(a)) reveals the progression of dynamical states as F is varied. At higher amplitudes (around F≈4.0), the system exhibits a stable period-1 orbit, shown by a single continuous branch in the diagram. A transition to a period-2 orbit occurs as the amplitude decreases, indicating a bifurcation point. Interestingly, a reappearance of the period-1 orbit is observed in the range F=3.0 to F=3.5, followed by further transitions to higher periodicities that suggest a cascade of period-doubling bifurcations.

At intermediate forcing amplitudes, the system shows intermittent chaos with periodic windows. For F≤0.25, persistent chaotic dynamics dominate. Phase portraits (Figures 1(b), 1(d), 1(f), 1(h), 1(j)) and time series (Figures 1(c), 1(e), 1(g), 1(i), 1(k)) mainly exhibit period-1 orbits, except at F=0.05, where a thickened limit cycle indicates chaotic behavior.

In the bifurcation diagram in Figure 2(a), when μ = 2 and ω = 1, increasing F drives the system from chaotic to period-1 behavior, evidenced by a single continuous branch. Compared to Figure 1, this diagram exhibits a stable period-1 orbit more clearly, with fixed-point values oscillating near zero. Between F=2.5 and F=3.0, a period-2 orbit emerges, indicated by bifurcation into two distinct branches. Although the diagram is dominated by period-1 orbit, regions of chaos appear intermittently, characterised by scattered points reflecting the interplay between periodic and chaotic states. The chaotic behavior occurs at F=0.5, as shown by the complex loops in the phase portrait and irregular periodicities in the time series (Figures (d) and (e)). At F=0.05 (Figures (b) and (c)), a weaker form of chaos is observed. Outside these regions, the system maintains stable period-1 dynamics.

The bifurcation diagram (Figure 3(a)) reveals a dominant period-2 orbit, absent in Figures 1 and 2 for μ=2 and ω=3, as indicated by two distinct branches. At low forcing (F<2), the period-2 orbit appears densely packed, with the system exhibiting simple periodic behavior characterized by a single fixed point for each F. As F approaches zero, the chaotic orbit narrows. Beyond F≈2, successive bifurcations give rise to increasingly complex dynamics, with pronounced chaotic regions between 2<F<4 marked by scattered points. For F>4, the system regains periodicity with more structured branches, and for F>8, fewer distinct branches indicate the emergence of simpler periodic attractors under strong forcing. At F=10, a period-4 orbit appears, comprising two stable and two unstable branches. The symmetry of fixed points about x=0 reflects the inherent symmetry of the Van der Pol oscillator.

The phase portrait and time series at F=0.5 (Figures (b) and (c)) show a thick trajectory, indicating chaos within a period-1 orbit. At F=3.0 and F=6.0 (Figures (c) and (d)), the chaotic features diminish, consistent with the bifurcation structure. Finally, at F=9.5 (Figure (e)), the system stabilizes into a periodic attractor, confirming the re-emergence of ordered dynamics.

The bifurcation diagram in Figure 4(a) for μ=−0.5 and ω=1.0 demonstrates that, as F increases, the system predominantly maintains a period-1 orbit, reflected by a continuous line formed by sampled values of the state variable x. The phase portrait and time series at F=0.2 (Figure (b)) further confirm the existence of a distinct limit cycle, characteristic of period-1 dynamics. At higher forcing values (F=0.5,1.0,1.5), the phase portraits and time series indicate the emergence of fixed points from the period-1 orbit observed in the bifurcation diagram.

We analyze the system dynamics through bifurcation diagrams by varying the forcing amplitude F under two conditions: (i) integer and (ii) non-integer ratios between the applied and natural frequencies.

**(i) Integer Condition:** Figure 5 presents the bifurcation diagram for the forced Van der Pol oscillator with an integer frequency ratio of 2:1. For low forcing (F<0.5), the system exhibits period-2 behavior, evident from two distinct branches. As F increases to the intermediate range (0.5≤F≤1.0), bifurcations occur, leading to the emergence of multiple branches indicative of complex periodic or quasi-periodic dynamics. At higher forcing amplitudes (F>1.0), the diagram displays densely packed and overlapping branches, characteristic of chaotic or multi-periodic behavior. Notably, for F>1.8, the strong merging of branches signifies a transition to chaos, marked by sensitive dependence on initial conditions.

Phase portraits at F=0.25 and F=0.75 (Figures 5(b) and 5(d)) reflect this transition between attractors. At F=0.25, trajectories are periodic, consistent with stable limit cycle behavior. In contrast, at F=0.75, the phase portrait exhibits a proliferation of trajectories, indicating the onset of chaotic dynamics as the forcing amplitude increases.

**(ii) Non-integer Condition:** Figure 6 depicts the bifurcation diagram for the forced Van der Pol oscillator with a non-integer frequency ratio of 5:3. Compared to the integer case (Figure 5), the system exhibits a wider chaotic behavior in the parameter space. At F=0.25, the bifurcation diagram shows a single continuous branch corresponding to a period-1 orbit with a narrow band chaos. At F=1.75, a period-2 orbit emerges, marked by two distinct branches, one significantly denser, indicating the probability of occurrence of the branches. Phase portraits and time series (Figures 6(b) and 6(c)) further illustrate this behavior: at F=0.25, the system converges to a stable limit cycle with a little irregularities, while at higher F, particularly in Figure 6(c), the system displays fully developed chaos without a well-defined limit cycle, consistent with the bifurcation structure.

Figure 7(a) shows that for F<0.75, the system exhibits a predominantly period-1 orbit with slight chaos, indicated by a continuous line with scattered points. In the range 0.75<F<1.5, the system alternates between period-1 and period-2 orbits accompanied by chaotic behavior. At higher F values, the dynamics transition to fully aperiodic, i.e., a chaotic motion, as reflected by the dense scattering of points in the bifurcation diagram.

The corresponding phase portraits and time series confirm this progression: at F=0.5, the system remains close to a stable limit cycle; at F=1.0, chaotic features become more prominent; and by F=1.5, the system exhibits strongly aperiodic, chaotic behavior. This trend emphasizes the role of increasing forcing amplitude in driving the system toward enhanced chaos.

The bifurcation diagram in Figure 8(a) reveals the evolution of chaotic behavior with varying F for the ratio 2.494/1. As F increases, the size of the chaotic attractor expands. Initially, for F<0.85, the system exhibits a period-1 orbit, which transitions to a period-2 orbit for 0.85<F<1.0 and then to a period-4 orbit for 1.0<F<1.15. Between F=1.0 and F=1.25, a periodic window is observed. Beyond this range, the system alternates between period-4 and period-3 orbits before entering fully chaotic dynamics. Thus, the system displays intermittency, transitioning from chaos at low F to periodicity at intermediate F, and back to chaos at higher F.

The phase portraits support this behavior: at F=1.2, the system already shows substantial chaotic features with multiple distinct loops, while at F=1.75, the chaotic attractor becomes wider and less structured, indicating stronger divergence and the absence of a stable limit cycle compared to earlier figures (Figures 1–4).

The above discussion highlights that the system exhibits a broad range of bifurcation behaviors, transitioning between periodic and chaotic dynamics depending on the ratio values. In a typical continuous system, period-doubling bifurcations are observed; however, with the application of the forcing function, the system demonstrates more complex dynamics. Thus, nonlinear oscillators subjected to forcing display irregular behaviors, emphasizing the intricate interplay between periodic and chaotic motion.

**Figures and Figure Captions**

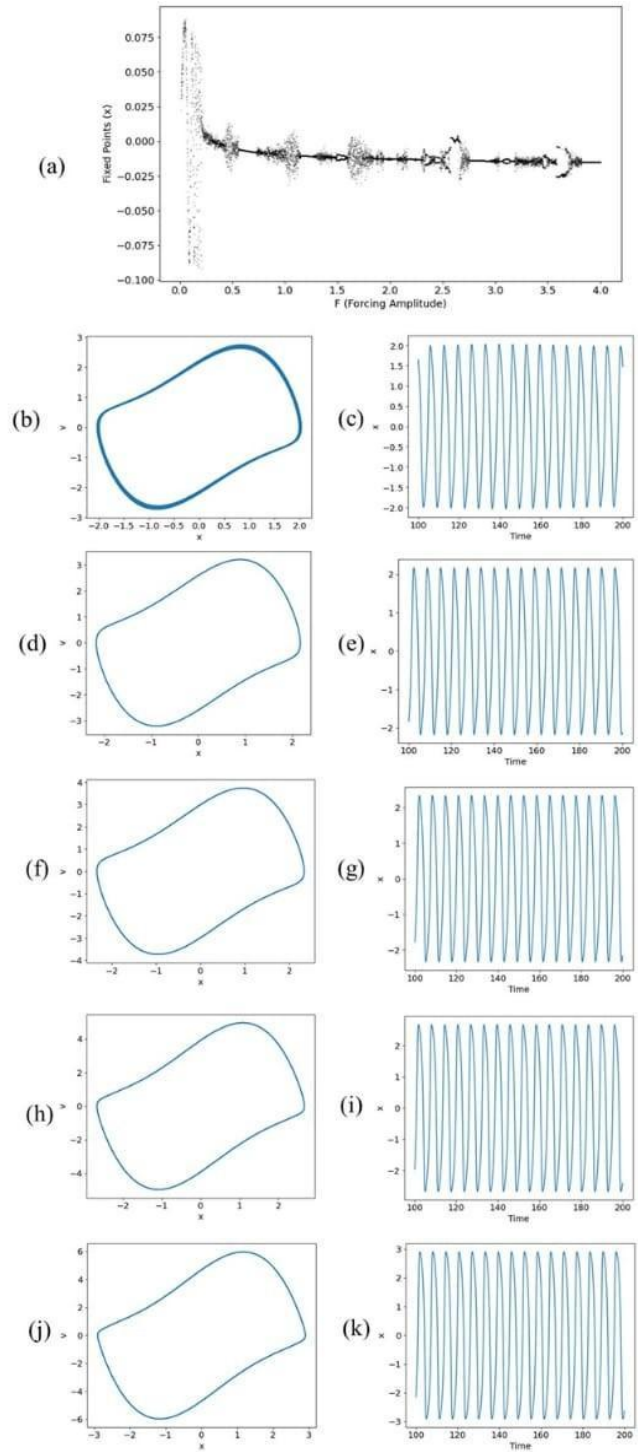

**Figure 1: Bifurcation and dynamical behavior of the forced Van der Pol oscillator at μ=1, ω=1, and varying forcing amplitudes F.** (a) Bifurcation diagram showing the evolution of system dynamics as F varies from 0 to 4. (b, d, f, h, j) Phase portraits of the oscillator at F=0.05, 0.5, 1.0, 2.5, and 4.0, respectively. (c, e, g, i, k) Corresponding time series plots for the same values of F. The state of the oscillator was sampled stroboscopically and synchronized with the external periodic forcing. All simulations were performed numerically using a fixed time step and standard integration methods.

These plots illustrate transitions from periodic to increasingly complex dynamical behavior as the forcing amplitude increases.

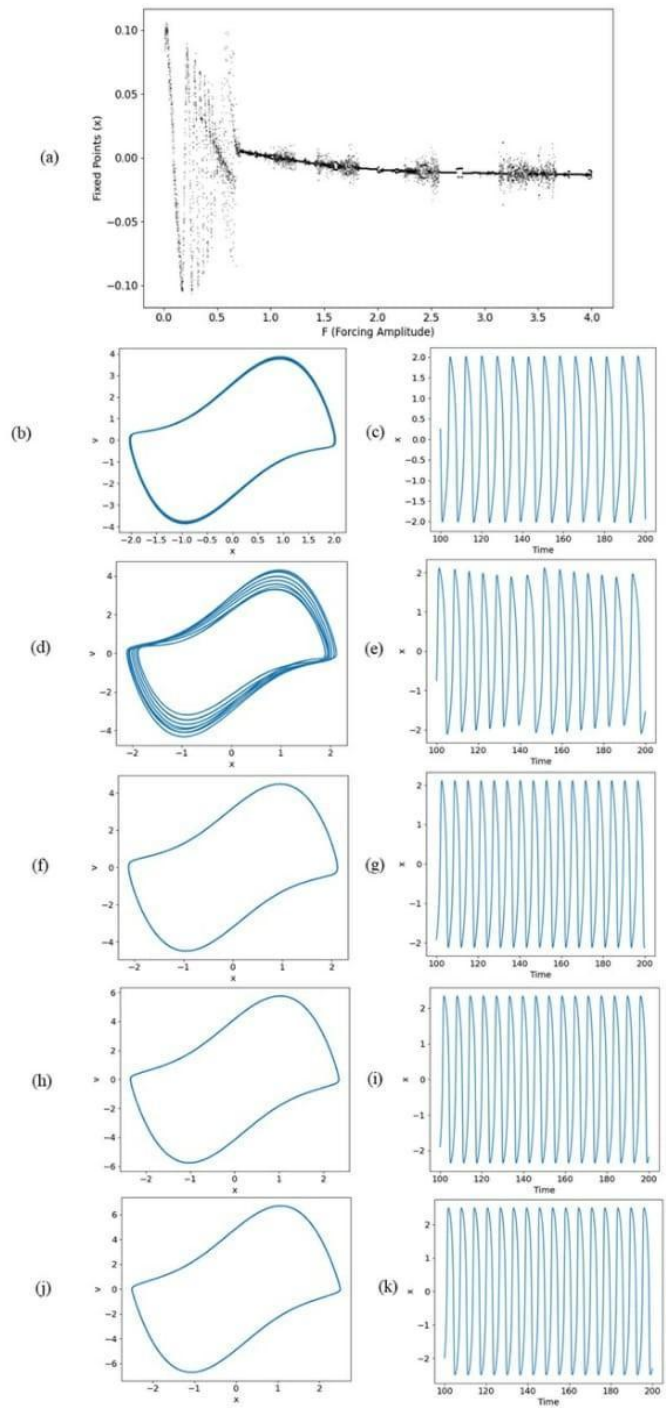

**Figure 2: Dynamical response of the forced Van der Pol oscillator at μ=2, ω=1, and varying forcing amplitudes.** (a) Bifurcation diagram showing the evolution of the system as the forcing amplitude F varies from 0 to 4. Phase portraits in (b, d, f, h, j) and corresponding time series in (c, e, g, i, k) are shown for F=0.05, 0.5, 1.0, 2.5, and 4.0, respectively. Simulations were conducted using

numerical integration with fixed time steps, and state variables were sampled stroboscopically. These results reveal transitions from periodic to chaotic behavior as F changes.

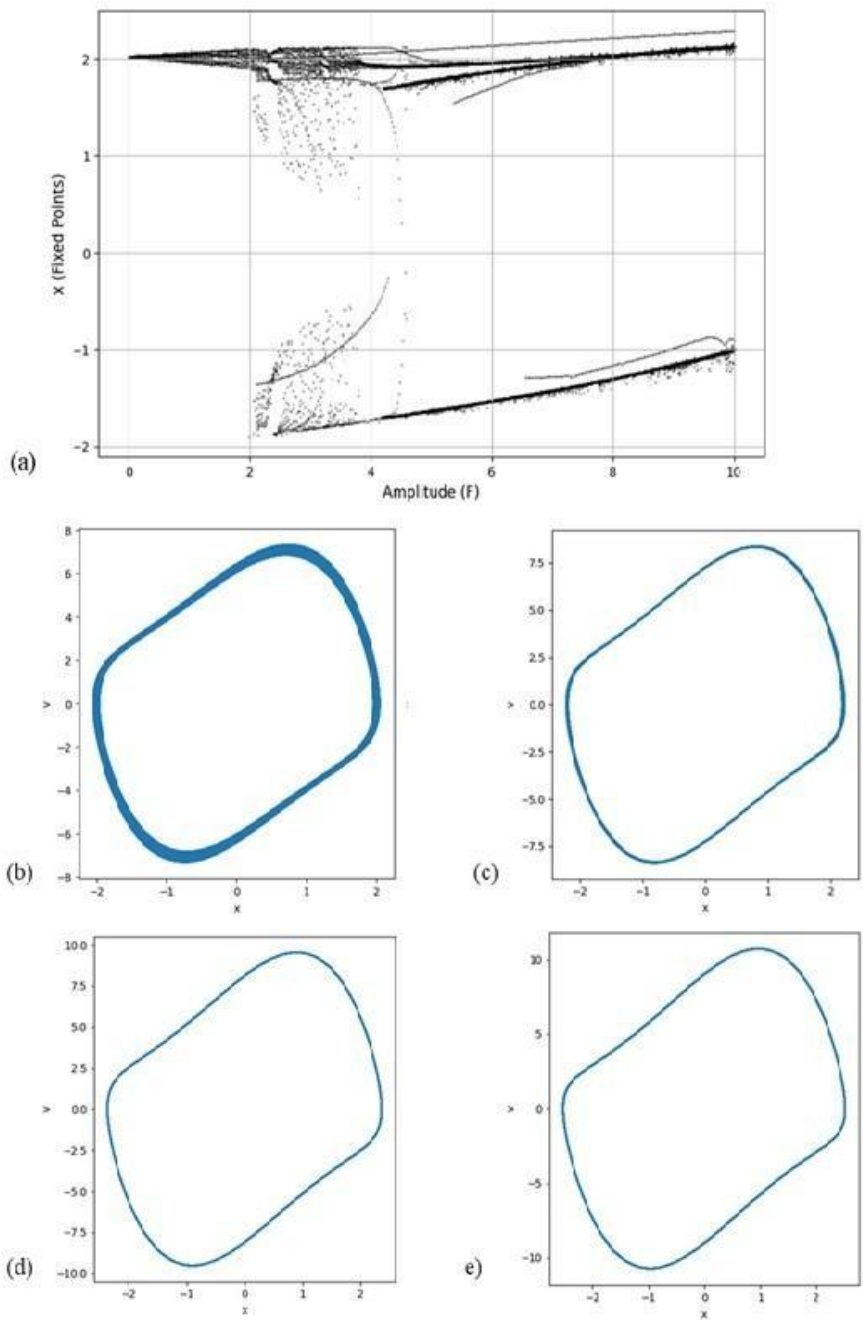

**Figure 3: Bifurcation and phase space dynamics of the forced Van der Pol oscillator at µ=2 and ω=3.** (a) Bifurcation diagram showing the system's response as the forcing amplitude F increases from 0 to 4. Phase portraits in (b–e) illustrate the oscillator's behavior at F=0.5, 3.0, 6.0, and 9.5, respectively. Results highlight the emergence of complex dynamics under higher frequency forcing conditions.

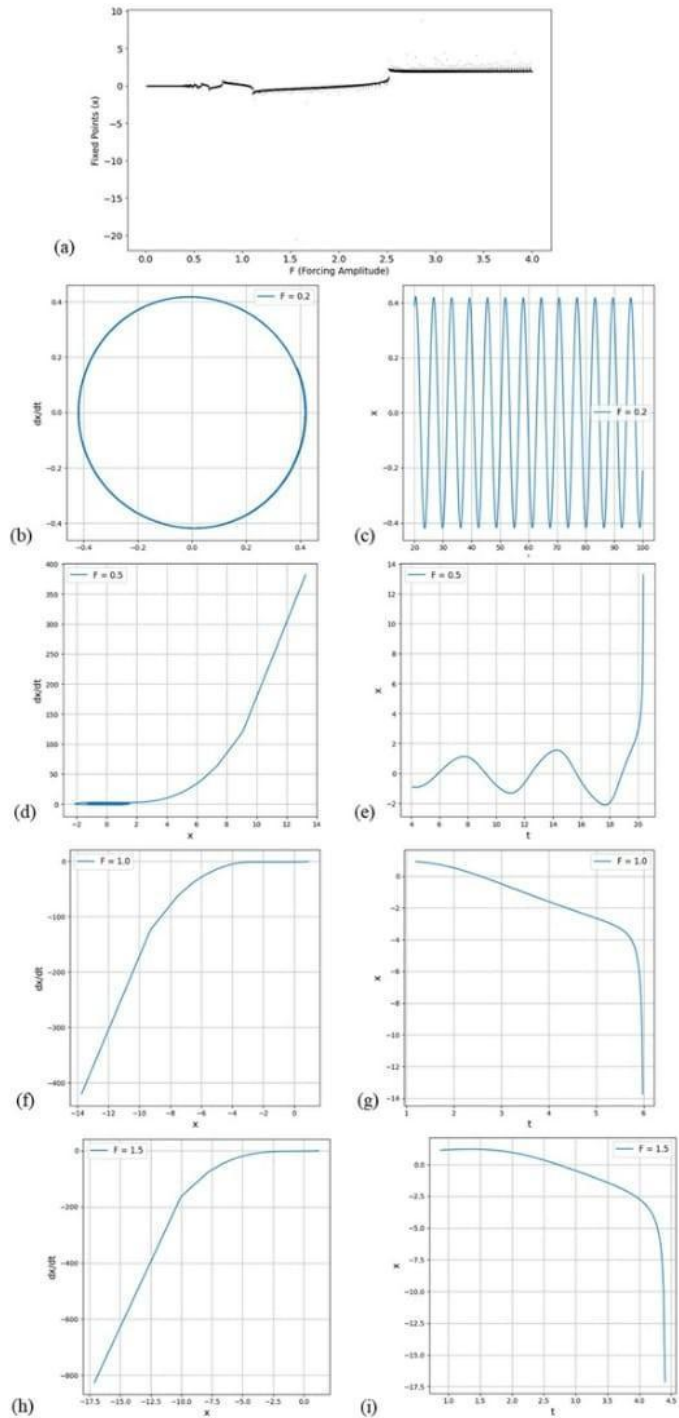

**Figure 4: Dynamics of the forced Van der Pol oscillator at μ=−0.5, ω=1.0, under varying forcing amplitudes.** (a) Bifurcation diagram showing system behavior as forcing amplitude F varies from 0 to 4. Phase portraits in (b, d, f, h) and corresponding time series in (c, e, g, i) are shown for F = 0.2, 0.5, 1.0, and 1.5, respectively. The results reveal the influence of negative damping on the transition between periodic and aperiodic dynamics.

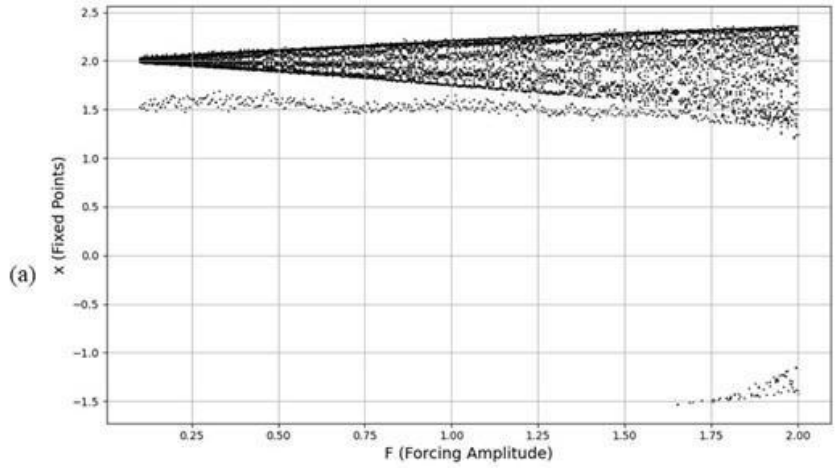

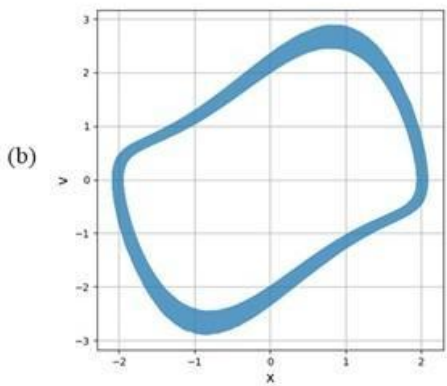
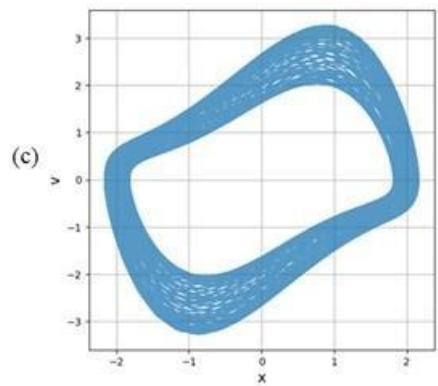

**Figure 5: Influence of frequency ratio 2:1 on the dynamics of the forced van der Pol oscillator at µ=1.** (a) Bifurcation diagram showing the variation in system dynamics as the forcing amplitude F increases from 0 to 4. Phase portraits in (b) and (d), and corresponding time series in (c) and (e), illustrate the response of the oscillator at F=0.25 and F=0.75, respectively. Simulations were performed for a frequency ratio 2:1 between the applied and natural frequencies. Each dataset represents a single trial.

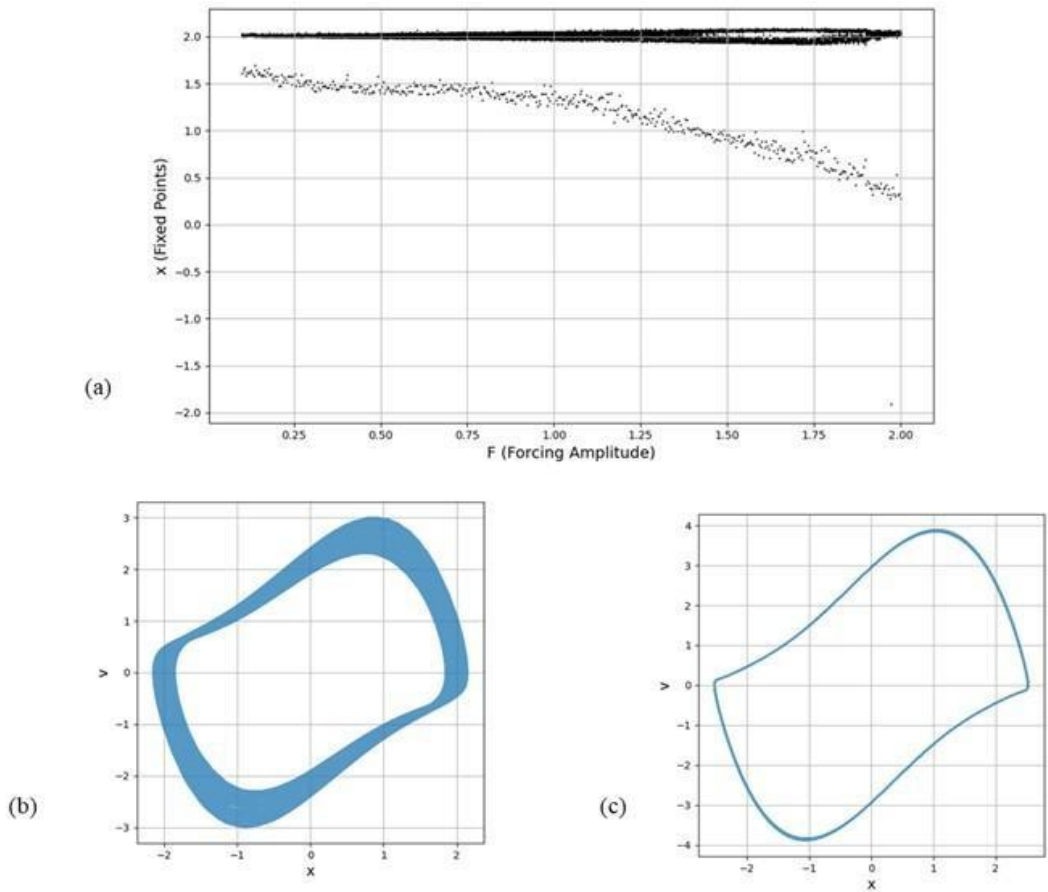

**Figure 6: Effect of frequency ratio 5:3 on the dynamics of the forced van der Pol oscillator at µ=1.** (a) Bifurcation diagram illustrating changes in system behavior as the forcing amplitude F varies from 0 to 4. Phase portraits in (b) and (c) correspond to F=0.25 and F=1.75, respectively, with a 5:3 ratio applied to natural frequency. Each simulation represents a single trial.

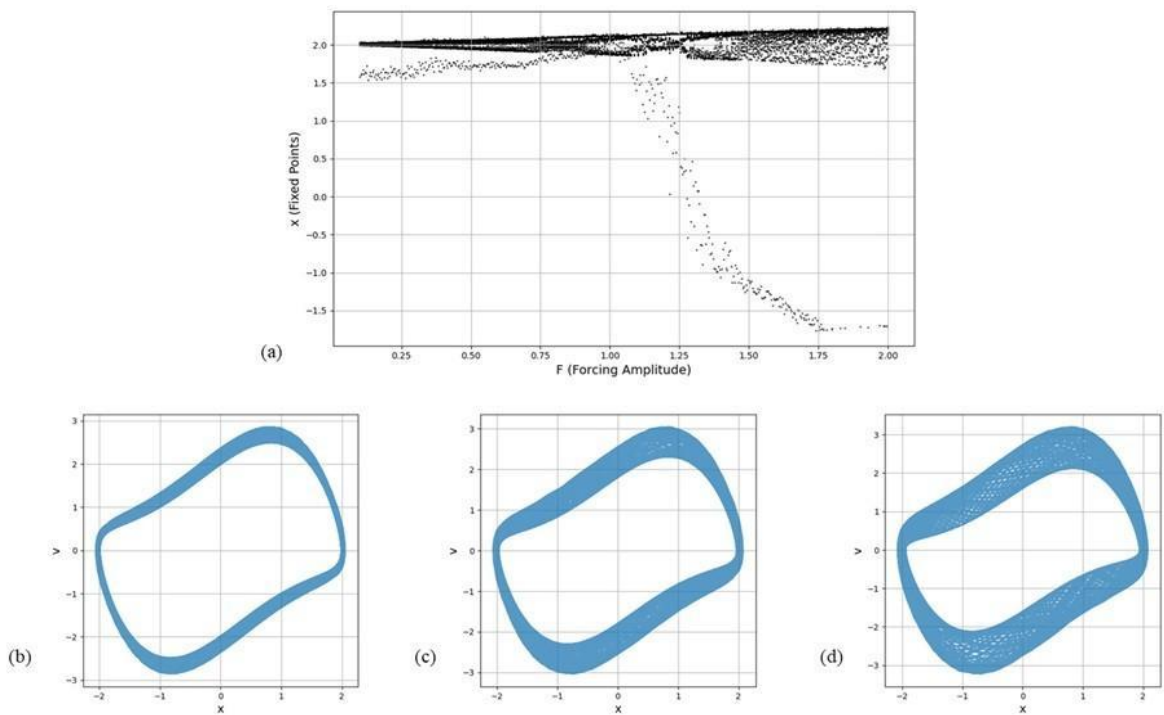

**Figure 7: Influence of frequency ratio 10:3 on the dynamics of the forced van der Pol oscillator at μ=1.** (a) Bifurcation diagram showing system behavior as the forcing amplitude F varies from 0 to 4. Phase portraits in (b), (c), and (d) correspond to F=0.5, 1.0, and 1.5, respectively, with an applied-to-natural frequency ratio of 10:3. Each dataset is based on a single simulation.

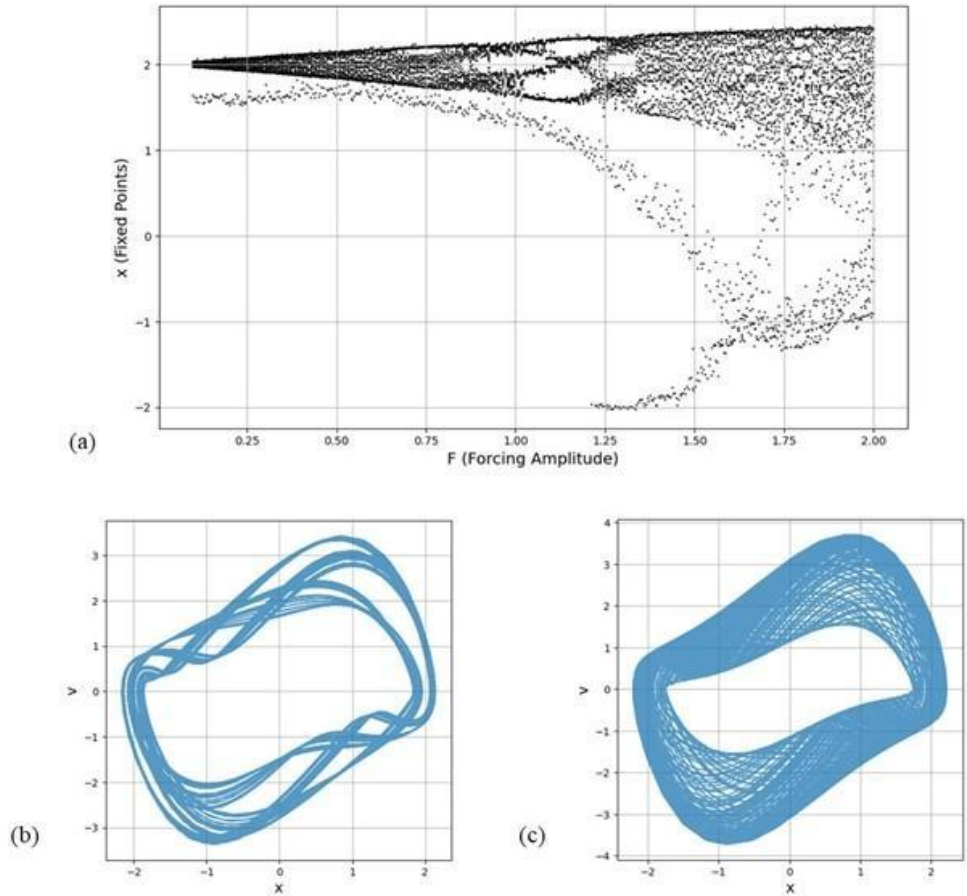

**Figure 8: Effect of near-resonant frequency ratio (2.494:1) on the dynamics of the forced van der Pol oscillator at μ=1.** (a) Bifurcation diagram illustrating system response as forcing amplitude F varies from 0 to 4. Phase portraits in (b) and (c) correspond to F=1.2 and 1.75, respectively, with an applied-to-natural frequency ratio of 2.494:1. Each simulation reflects one set of initial conditions.

## DISCUSSION

The forced Van der Pol oscillator exhibits nonlinear behaviors such as period-doubling and chaos, which are relevant to real-world systems in astrophysics, biology, and engineering. In celestial mechanics, similar transitions from periodic to chaotic orbits occur in resonant planetary systems due to gravitational forcing, akin to the bifurcation patterns observed in Figures 1–3 (Murray and Dermott [16]). In biology, the oscillator models phenomena like cardiac arrhythmias and neuronal firing, where increased stimulus leads to transitions from regular to chaotic activity. These dynamics, illustrated in

Figures 1, 2, and 4, mirror period-doubling bifurcations in heart rhythms and chaotic spiking in neurons (Glass and Mackey [17]).

In engineering, the periodic and chaotic behaviors of the forced Van der Pol oscillator are pertinent to nonlinear dynamics in electrical circuits and power systems. Relaxation oscillators and converters often exhibit bifurcations and transitions to chaos under varying input conditions, as reflected in Figures 5–7. In particular, increasing the forcing amplitude leads to a shift from stable periodic to chaotic behavior, analogous to voltage instabilities observed in power grids. These dynamics highlight the sensitive dependence on initial conditions and parameter changes, where multiple coexisting states and irregular oscillations emerge under strong forcing, paralleling instabilities in real-world electrical systems.

This study examines the dynamical behavior of the forced Van der Pol oscillator under varying forcing amplitudes, damping coefficients, and driving frequencies. The system exhibits a rich spectrum of responses, as evidenced by bifurcation diagrams, phase portraits, and time series. At lower forcing amplitudes, the oscillator often displays chaotic dynamics, characterized by irregular trajectories and scattered bifurcation branches. As the forcing amplitude increases, the system transitions to periodic behavior, including stable period-1 and period-2 orbits. Intermediate forcing levels sometimes led to intermittent chaos, where periodic and chaotic phases alternated. At higher amplitudes, the dynamics generally stabilizes, with dominant low-period attractors and diminished chaotic behavior. These results underscore the sensitivity to parameter variations of the system and the emergence of complex transitions between stable and unstable regimes.

The dynamics of the forced Van der Pol oscillator are also found to depend strongly on the ratio of the applied frequency to the natural frequency of the system. Bifurcation diagrams, phase portraits, and time series reveal that for integer frequency ratios, the system exhibits a progression from periodic to chaotic behavior with increasing forcing amplitude. The period-2 orbits dominate at low amplitudes. As the amplitude increases further, the system undergoes bifurcations, leading to higher periodicities to quasi-periodic-like dynamics. At higher amplitudes, densely overlapping branches emerge in the bifurcation diagrams, indicating chaotic or multi-periodic regimes.

For non-integer frequency ratios, the oscillator exhibits a shift from stable period-1 orbits at low forcing amplitudes to higher periodicities and chaotic behavior at higher amplitudes. Phase portraits show the absence of stable limit cycles, especially under irrational ratios, where intermittent dynamics emerged. Periodic windows (e.g., period 2, 3, and 4) are interspersed within chaotic regimes, with intermediate amplitudes revealing transient order before transitioning to fully aperiodic motion.

The dynamics of the forced Van der Pol oscillator strongly depend on the ratio of applied to natural frequencies. Integer ratios yield both periodic and chaotic behavior, with chaos emerging at higher forcing amplitudes. Non-integer and irrational ratios produce complex dynamics, featuring intermittent transitions across broader amplitude ranges. As forcing increases, the response of the system grows more sensitive and complex. While the unforced oscillator remains regular, forcing introduces irregularities, making the system suitable for modeling phenomena in astrophysics, biology (e.g., cardiac arrhythmias), and engineering (e.g., nonlinear circuits).

**MATERIALS AND METHODS**

In this study, we investigated the nonlinear dynamics of the forced van der Pol oscillator by analyzing its behavior across a range of forcing amplitudes. The system was modeled using a second-order ordinary differential equation (ODE) that incorporates nonlinear damping and an external periodic forcing term. The mathematical form of the equation used is $\frac{d^2x}{dt^2} + \mu(x^2 - 1)\frac{dx}{dt} + x = F\cos(\omega t)$, where μ is the nonlinearity parameter, F is the forcing amplitude, and ω is the forcing frequency.

The numerical simulations were conducted in Python using the scipy.integrate.odeint routine from the SciPy library to solve the coupled system of first-order ODEs. The function odeint was employed to integrate the system with specified initial conditions and parameter values. Specifically, the dynamics were explored over a time range of t=0 to 50 seconds, sampled uniformly with 5000 points. The initial conditions for the state variables were set as x(0)=2 and $\frac{dx(0)}{dt}$=0.

For each simulation, the parameters μ=1.0, β=1.0, and ω=2.0 were held constant while the forcing amplitude F was varied to examine transitions between periodic, quasi-periodic, and chaotic behavior. The time series were obtained by plotting x(t) as a function of time, and phase portraits were generated by plotting x(t) against $\frac{dx}{dt}$, revealing the nature of the attractors of the system.

To construct bifurcation diagrams, the forcing amplitude F was incrementally varied over a defined range (e.g., 0.1 to 2.0) in 200 steps. For each value of F, the ODE was numerically solved, and the final segment of the solution (last 100 time steps) was extracted to identify the steady-state behavior of the system. These values were plotted against their respective F values to visualize bifurcation structures and identify transitions between periodic and chaotic regimes.

All plots were generated using matplotlib.pyplot (Matplotlib library in Python) and NumPy were used for numerical computations and array manipulations. The analysis enabled the classification of system behavior into particular dynamical regimes, supported by time series, phase portraits, and bifurcation diagrams.

**ACKNOWLEDGEMENTS**

No acknowledgement.

**APPENDIX**

The codes have been generated in Python. The codes are available at the link:
https://github.com/Astriek/bifurcation/blob/main/bifrucation.py